\documentclass[pra,twocolumn,showpacs]{revtex4}
\usepackage{graphicx,graphics,psfrag,amsmath,calc,amssymb}

\begin{document}

\title{Finite temperature phase diagram of trapped Fermi gases with population
imbalance}
\author{Wei Zhang and L.-M. Duan}
\affiliation {FOCUS center and MCTP, Department of Physics,
University of Michigan, Ann Arbor, MI 48109}
\date{\today}

\begin{abstract}
We consider a trapped Fermi gas with population imbalance at finite
temperatures and map out the detailed phase diagram across a wide Feshbach
resonance. We take the Larkin-Ovchinnikov-Fulde-Ferrel (LOFF) state into
consideration and minimize the thermodynamical potential to ensure
stability. Under the local density approximation, we conclude that a stable
LOFF state is present only on the BCS side of the Feshbach resonance, but
not on the BEC side or at unitarity. Furthermore, even on the BCS side, a
LOFF state is restricted at low temperatures and in a small region of the
trap, which makes a direct observation of LOFF state a challenging task.
\end{abstract}

\pacs{03.75.Ss, 03.75.Hh, 05.30.Fk}

\maketitle

%
%

\section{Introduction}

\label{sec:introduction}

There has been considerable interest in paired
superfluidity of trapped Fermi gases, where the interatomic
interaction can be tuned by varying the external magnetic
field~\cite{regal-04,zwierlein-04,chin-04}. Recently, the
experimental realization of superfluidity in polarized Fermi
gases attracts great attention, where the numbers $N_{\uparrow }$
and $N_{\downarrow }$ of the two atomic species undergoing pairing
are different~\cite{zwierlein-06,partridge-06,zwierlein-06b}.
This population imbalance is obviously detrimental to superfluidity,
since the Cooper pairing requires equal number of atoms from both
spin components. Therefore, by increasing the population imbalance from zero,
it is expected that the BCS pairing state becomes less favorable and
eventually gives its way to the normal or other exotic
phases~\cite{LOFF, sarma-63, sedrakian-02, liu-03, bedaque-03,wei1}.

One of the most fascinating phenomena in unbalanced Fermi systems is the
Larkin-Ovchinnikov-Fulde-Ferrel (LOFF) states, which was first discussed as
a "compromise" candidate exhibiting both pairing and non-zero magnetization
in the context of superconductors in the presence of a magnetic field~\cite{LOFF}.
This exotic LOFF state is characterized by an order parameter with
one or more non-zero components at finite momenta $\mathbf{q}$, hence
breaks translational and rotational symmetry and forms a crystal of pairing
order (i.e., a supersolid). In the past several decades, the existence of
LOFF states was studied in various systems~\cite{casalbuoni-rev}, including
heavy fermions~\cite{radovan-03} and dense quark matter~\cite{alford-01}.

Comparing to the systems mentioned above, ultra-cold Fermi gases
provide a super clean experimental platform with remarkable
controllability, so that they can be studied with nearly arbitrary
interaction strength and population imbalance. Therefore, after
the realization of paired superfluidity in resonantly interacting
$^{6}$Li atoms with population imbalance~\cite{zwierlein-06,partridge-06},
the interest of LOFF state in these systems has been greatly intensified.
Up to now, no evidence of LOFF state has been found yet in
experiments on the polarized Fermi gas. Theoretically, some existing
studies give significantly different predictions on the stable region of
LOFF state: some works on the homogeneous system conclude
that at zero temperature the LOFF state is confined to a narrow parameter region
on the BCS side of the resonance, for both cases of a narrow~\cite{sheehy-06}
and a wide \cite{wei2} Feshbach resonance; while some other calculations
claim a much larger region of a stable LOFF state~\cite{machida-06,kinnunen-06, he-07},
well existing at the unitarity point. What is subtle
for these calculations is to implement a right set of
stability conditions, which are usually controversial.
Considering that there exist various possible competing
phase configurations for this system, one thus needs to carefully
distinguish a stable phase from some metastable states or
unstable saddle point solutions.

In this manuscript, we consider the possibility of LOFF state
and map out the detailed finite temperature phase diagram for
fermionic atoms in a trap with population imbalance. To ensure
stability, we directly minimize the thermodynamical potential
instead of using the order parameter equations, as the latter may
give unstable or metastable solutions~\cite{wei3}. The calculation
of the full landscape of thermodynamical potential sounds to
us the only method capable to distinguish a local metastable
configuration from a globally stable phase.
To deal with the trap, we use the local density approximation (LDA)
which is typically valid unless the trap has a strong anisotropy
and/or the total atom number is small~\cite{partridge-06,muller,note1}.
Our main results are as follows. (i) We conclude that a stable LOFF state
can only be present on the BCS side of the Feshbach resonance. On
the BEC side and at the unitarity, the LOFF state is only a
metastable state. (ii) Even on the BCS side of the resonance,
a globally stable LOFF state is only restricted at low temperatures
and in a small region of the trap. In a wider temperature
and spatial region, the LOFF state is only a metastable state.
Considering experimental limit of temperature and resolution, we
expect that a direct observation of stable LOFF states is challenging
for polarized Fermi gases in typical harmonic traps.

The remainder of this manuscript is organized as follows. In
Section~\ref{sec:formalism}, we discuss our formalism for unbalanced
Fermi gases in an isotropic harmonic trap. We first derive a
Ginzburg--Landau theory for the second order normal-LOFF and normal-BCS
phase transitions, and then derive the mean-field thermodynamical
potential to study the first order normal-LOFF and LOFF-BCS phase transitions.
In the mean-field calculation, for simplicity, we focus on the
single-plane-wave LOFF state (i.e., the FF state).
Our main results are presented in Section~\ref{sec:phasediagram},
where the phase diagrams showing various stable or metastable states
are illustrated for systems at unitarity, and on the BCS and BEC sides
of the Feshbach resonance.


\section{The formalism for polarized Fermi gases in a harmonic trap}

\label{sec:formalism}

Since the population of closed channel molecules is negligible close to
a wide Feshbach resonance~\cite{chen-05,yi-06a}, we study a trapped Fermi gas
by considering the following single-channel Hamiltonian (we use the natural
unit such that $\hbar=k_{B}=1$ throughout this manuscript):
\begin{eqnarray}
\mathcal{H} &=& \sum_{\mathbf{k},\sigma } \xi _{\mathbf{k},\sigma }
a_{\mathbf{k},\sigma }^{\dagger } a_{\mathbf{k},\sigma }
\nonumber \\
&&\hspace{-1cm}
+\sum_{\mathbf{k},\mathbf{k}^{\prime },\mathbf{q}}
V(\mathbf{k},\mathbf{k}^{\prime })
a_{\mathbf{q}/2+\mathbf{k},\uparrow }^{\dagger }
a_{\mathbf{q}/2-\mathbf{k},\downarrow }^{\dagger }
a_{\mathbf{q}/2-\mathbf{k}^{\prime },\downarrow }
a_{\mathbf{q}/2+\mathbf{k}^{\prime },\uparrow },
\label{eqn:hamiltonian}
\end{eqnarray}
where $a_{\mathbf{k},\sigma }^{\dagger }$ and $a_{\mathbf{k},\sigma }$ are
creation and annihilation operators for fermions labeled by the spin
(hyperfine state) indices $\sigma =\uparrow ,\downarrow $, respectively, and
$\xi_{\mathbf{k},\sigma }=\epsilon_{\mathbf{k}}-\mu_{\sigma }$ is the
fermion dispersion $\epsilon_{\mathbf{k}}=k^{2}/(2m)$ shifted by the
corresponding chemical potential. The attractive fermion-fermion interaction
$V(\mathbf{k},\mathbf{k}^{\prime })$ can be written in a BCS form as
$V(\mathbf{k},\mathbf{k}^{\prime })=-g$, provided that only $s$-wave
contact interaction is considered. The interaction strength $g$ can be
connected with scattering parameters through the standard renormalization relation
\begin{equation}
-\frac{1}{g}=\frac{\mathcal{N}_{0}}{k_{F}a_{s}}-\sum_{\mathbf{k}}\frac{1}{%
2\epsilon _{\mathbf{k}}},  \label{eqn:renormalization}
\end{equation}
where $\mathcal{N}_{0}=mL^{3}k_{F}/(4\pi )$ with Fermi momentum $k_{F}$
and quantization volume $L^{3}$, and $a_{s}$ is the $s$-wave
scattering length. In the following discussion, we take the local density
approximation such that $\mu _{\uparrow }=\mu (\mathbf{r})+h$, $\mu
_{\downarrow }=\mu (\mathbf{r})-h$, and $\mu (\mathbf{r})=\mu_{0}-U(\mathbf{r})$,
where $U(\mathbf{r})= m \omega ^{2}r^{2}/2$ represents an external
harmonic trap~\cite{note2}. The chemical potential at the trap
center $\mu_{0}$ and the chemical potential imbalance $h$ can be related to
the total particle number $N=N_{\uparrow }+N_{\downarrow }$ and the
population imbalance $P=(N_{\uparrow }-N_{\downarrow })/(N_{\uparrow}+N_{\downarrow })$.

Using the functional integral technique, we introduce the standard
Hubbard-Stratonovich field $\Delta $ which couples to $a^{\dagger}a^{\dagger }$
in order to integrate out the fermions, leading to the partition function
\begin{equation}
\mathcal{Z} = \mathrm{Tr}\left( e^{-\beta \mathcal{H}}\right)
=\int \mathcal{D}[\Delta ^{\dagger},\Delta ]
\exp \left\{ -S_{\mathrm{eff}}[\Delta ^{\dagger },\Delta ]\right\}
\label{eqn:partition}
\end{equation}
with the effective action
\begin{equation}
S_{\mathrm{eff}}[\Delta ^{\dagger },\Delta ]
=\int_{0}^{\beta }\frac{d\tau }{\beta }
\sum_{\mathbf{k}}\left\{ \frac{\beta |\Delta |^{2}}{g}
-\mathrm{Tr}\ln \left( \beta \mathbf{G}^{-1}[\Delta ]\right) \right\} ,
\label{eqn:action}
\end{equation}
where $\Delta \equiv \Delta (\mathbf{q},\tau )$ depends on momentum $\mathbf{q}$
and imaginary time $\tau $, $\beta =1/T$ is the inverse temperature, and
$\mathbf{G}^{-1}$ is the inverse fermion propagator
\begin{equation}
\mathbf{G}^{-1}[\Delta (\mathbf{q},\tau )]=
\left(
\begin{array}{cc}
-\partial_{\tau }+\xi _{\uparrow } & \Delta (\mathbf{q},\tau ) \\
\Delta ^{\dagger }(\mathbf{q},\tau ) & -\partial_{\tau }-\xi_{\downarrow }
\end{array}
\right).  \label{eqn:propagator}
\end{equation}

Up to now we have not incorporated any approximation and the effective
action Eq. (\ref{eqn:action}) is accurate. Next, we will discuss in the rest
of this section two approximation schemes, which can be applied to various
situations. In section~\ref{sec:GLtheory}, a Ginzburg--Landau (GL) theory is
presented to study the second order normal-LOFF and normal-BCS phase
transitions. In section~\ref{sec:MFtheory}, a mean-field approach is applied
to derive thermodynamical potential, such that the possibilities of first
order normal-BCS and LOFF-BCS phase transitions can be analyzed.


\subsection{Ginzburg--Landau theory and 2nd order phase transitions}

\label{sec:GLtheory}

\begin{figure}[tbp]
\begin{center}
\psfrag{r}{$r$}
\psfrag{mu}{$\mu$}
\includegraphics[width=8.0cm]{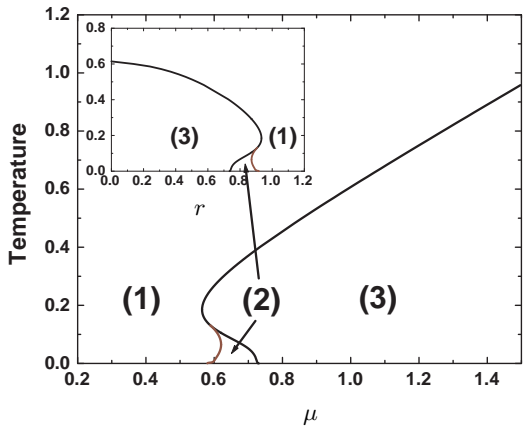}
\end{center}
\caption{ Temperature versus chemical potential phase diagram is illustrated
by solving Eq. (\ref{eqn:GLequation}) at unitarity, showing three different
regions: (1) normal state is stable or metastable, (2) normal state loses
its stability at $\mathbf{q}\neq 0$, and (3) normal state loses its
stability at $\mathbf{q}=0$. In this figure, the chemical potential
difference $h=0.4$ with arbitrary energy unit. Inset: The $T$-$r$ diagram can be
obtained by incorporating the LDA approximation. Here, we choose the energy
unit as the chemical potential at the trap center $\mu_{0}=1$,
and $h=0.4$ in this scale. The radius $r$ is in unit of
$r_0 \equiv \sqrt{2 \mu_0/(m \omega^2)}$.}
\label{fig:GL-Tc}
\end{figure}

In this section, we consider only the possibility of second order phase
transitions from normal to superfluid phase, including both ordinary BCS and
exotic LOFF states. Close to the phase transition line, the order
parameter $\Delta (\mathbf{q},\tau )$ has small magnitude, hence the
effective action $S_{\mathrm{eff}}$ can be expanded in powers of $\Delta $.
In the spirit of Ginzburg--Landau (GL) theory, we are interested in static
$\Delta (\mathbf{q})$. Therefore, the effective action takes the form
\begin{equation}
S_{\mathrm{eff}} \approx
\sum_{\mathbf{q}} \alpha (\mathbf{q})|\Delta (\mathbf{q})|^{2}
+\mathcal{O}(\Delta ^{4}),  \label{eqn:GLaction}
\end{equation}
where $\alpha (\mathbf{q})=g^{-1}-\chi (\mathbf{q},0)$ with
$\chi (\mathbf{q},0)$ is the pair susceptibility. Notice that as
$\alpha (\mathbf{q})$ depends only on the magnitude of wave vector $\mathbf{q}$,
the fourth order terms are crucial to determine the crystalline structures
of the LOFF state~\cite{casalbuoni-rev}. However, when we getting very close
to the phase transition line where the effective action is dominated by the
leading quadratic term, it is sufficient to consider the coefficient
$\alpha (\mathbf{q})$, leading to
\begin{eqnarray}
\alpha (\mathbf{q}) &=&
\sum_{\mathbf{k}} \left[ \frac{1}{2\epsilon _{\mathbf{k}}}
-\frac{1-n_{F}(\xi _{\mathbf{kq},+})-n_{F}(\xi _{\mathbf{kq},-})}
{2\xi _{\mathbf{kq}}}\right]
\nonumber  \label{eqn:quadratic-coefficient} \\
&&\hspace{1cm}-\frac{\mathcal{N}_{0}}{k_{F}a_{s}},
\end{eqnarray}
where $n_{F}(x) \equiv \left( e^{\beta x}+1\right) ^{-1}$ is the Fermi distribution,
$\xi_{\mathbf{kq},\pm }=\xi_{\mathbf{kq}}\pm (\delta\epsilon_{\mathbf{kq}}-h)$,
$\xi_{\mathbf{kq}}=\sum_{\sigma}(\xi_{\mathbf{k}+\mathbf{q},\sigma}+\xi_{\mathbf{k},\sigma})/4$,
and $\delta \epsilon _{\mathbf{kq}}=(\epsilon _{\mathbf{k}+\mathbf{q}}-\epsilon _{\mathbf{k}})/2$.
As concluded in the GL theory, the normal state with $\Delta (\mathbf{q})=0$
loses its stability as long as $\alpha (\mathbf{q})$ becomes negative for one
or more $\mathbf{q}$ components. Therefore, we introduce the order parameter equation,
which corresponds to the condition that $\alpha (\mathbf{q})$ crosses zero
\begin{equation}
\frac{\mathcal{N}_{0}}{k_{F}a_{s}}=\sum_{\mathbf{k}}
\left[ \frac{1}{2\epsilon _{\mathbf{k}}}
-\frac{1-n_{F}(\xi_{\mathbf{kq},+})-n_{F}(\xi_{\mathbf{kq},-})}
{2\xi_{\mathbf{kq}}}\right] .  \label{eqn:GLequation}
\end{equation}
This order parameter equation can be solved for a given $\mu (\mathbf{r})$
and $h$ to obtain the normal-BCS transition temperature $T_{c}(\mathbf{q}=0)$,
while the normal-LOFF transition temperature is determined by the maximal
$T_{c}(\mathbf{q})$ for all finite $\mathbf{q}\neq 0$. In Fig.~\ref{fig:GL-Tc},
a typical $T$-$\mu $ phase diagram is depicted showing three different
regions, where: (1) normal state is stable or metastable [$\alpha (\mathbf{q})>0$
for all $\mathbf{q}$], (2) normal state is unstable due to LOFF
instabilities [$\alpha (0)>0$, but $\alpha (\mathbf{q})<0$ for some finite
$\mathbf{q}$'s], and (3) normal state is unstable due to BCS instability
[$\alpha (0)<0$]. Notice that at unitarity, the scattering parameter
$a_{s}\rightarrow \infty $ and does not set up a length or energy scale.
Therefore, phase diagrams for different values of $h$ (as long as $h\neq 0$)
are identical when an appropriate energy unit is applied. Furthermore, by
incorporating the local density approximation (LDA)
$\mu (\mathbf{r})=\mu_{0}-U(\mathbf{r})$, this $T$-$\mu $ phase diagram
can be easily transformed into the $T$-$r$ plane ($r$ is the radius of the trap,
see inset of Fig.~\ref{fig:GL-Tc}), which can be related to experiments.

It should be emphasized that the phase diagrams obtained by solving the GL
equation $(\ref{eqn:GLequation})$ is not a result within mean-field~\cite{sheehy-06}
or NSR schemes~\cite{parish-07, hu-06}, since the according approximated number
equations are not included for self-consistent solving.
Therefore, under the assumption that the order parameter is small, the GL
theory discussed above is well controlled and the results are reliable.
However, the small order parameter restriction sets a limit of this approach
only for analyzing second order phase transitions. The complete phase
diagram, where the possibilities of first order phase transitions have to be
taken into consideration, is beyond the scope of GL theory since $|\Delta |$
is not necessarily small. Thus, we discuss in the next subsection a
mean-field approach to derive the thermodynamical potential, which offers an
approximation technique to study the complete phase diagram by direct
minimization.


\subsection{Mean-field theory and 1st order phase transitions}

\label{sec:MFtheory}

Unlike the GL expansion of effective action Eq. (\ref{eqn:action}) discussed
above, a mean-field theory is considered here by introducing a uniform
static saddle point $\Delta(\mathbf{q}, \tau) = \Delta_{\mathbf{Q}}
\delta(\mathbf{q}-\mathbf{Q})$, which corresponds to the BCS state for
$\mathbf{Q}=0$, and to the single-plane-wave LOFF state for finite $\mathbf{Q}$.
With this assumption, the saddle point action is
\begin{eqnarray}
\label{eqn:saddle-action}
S_0 &=& \beta \frac{\vert \Delta_{\mathbf{Q}} \vert^2}{g}
+ \sum_{\mathbf{k}}\Big\{ \beta \left( \xi_{\mathbf{kQ}} - E_{\mathbf{kQ}} \right)
\nonumber \\
&& +\ln\left[ n_F (-E_{\mathbf{kQ},+}) \right]
+ \ln\left[ n_F (-E_{\mathbf{kQ},-}) \right]\Big\},
\end{eqnarray}
where $E_{\mathbf{kQ}} = \sqrt{\xi^2_{\mathbf{kQ}} + \Delta_{\mathbf{Q}}^2}$,
and $E_{\mathbf{kQ}, \pm} = E_{\mathbf{kQ}} \pm (\delta \epsilon_{\mathbf{kQ}} - h)$
is the quasiparticle energy ($+$) or the negative of the quasihole energy ($-$).

The saddle point conditions $\delta S_{0}/\delta \Delta _{\mathbf{Q}}^{\dagger}=0$
and $\delta S_{0}/\delta |\mathbf{Q}|=0$ lead to the order parameter equations
\begin{eqnarray}
\frac{\mathcal{N}_{0}}{k_{F}a_{s}} &=&
\sum_{\mathbf{k}} \left[ \frac{1}{2\epsilon _{\mathbf{k}}}
-\frac{1-n_{F}(E_{\mathbf{kQ},+})-n_{F}(E_{\mathbf{kQ},-})}{2E_{\mathbf{kQ}}}\right],
\nonumber  \label{eqn:gap} \\
0 &=& \sum_{\mathbf{k}}(|\mathbf{Q}|-k_z)
\bigg\{ \left[ 1 + n_{F}(E_{\mathbf{kQ},+}) - n_{F}(E_{\mathbf{kQ},-}) \right]
\nonumber \\
&&\hspace{-5mm}+
\frac{\xi_{\mathbf{kQ}}}{E_{\mathbf{kQ}}}
\left[ -1 + n_{F}(E_{\mathbf{kQ},+}) + n_{F}(E_{\mathbf{kQ},-}) \right] \bigg\},
\end{eqnarray}
which can be solved together to determine $\Delta $ and $|\mathbf{Q}|$.
Here, $k_z = |\mathbf{k}| \cos\theta$ with $\theta$ is the polar angle.
However, solving these two order parameter equations is not sufficient to
determine the phase diagram, since the solutions may be metastable states or
only unstable saddle points. Therefore, to ensure stability, we evaluate
and directly minimize the thermodynamic potential $\Omega_{0}=S_{0}/\beta$,
instead of imposing various subtle stability criteria~\cite{he-07,sheehy-06}.
\begin{figure}[tbp]
\begin{center}
\psfrag{D}{$\Delta$}
\psfrag{Q}{$\vert {\bf Q} \vert$}
\includegraphics[width=8.0cm]{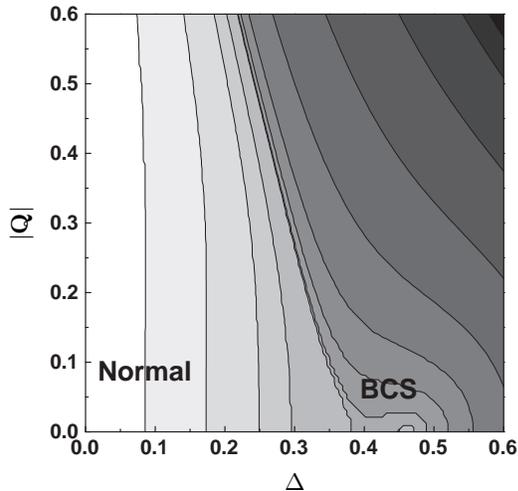}
\end{center}
\caption{Contour plots of mean-field thermodynamic potential $\Omega _{0}$
as a function of $\Delta$ and wave vector $|\mathbf{Q}|$. Two local
minima (lighter region) are shown to characterize the normal ($\Delta =0$)
and the BCS ($\Delta \neq 0$, $|\mathbf{Q}|$ =0) states. In this plot,
parameters are chosen as $h=0.4$, $\mu =0.5$, $T=0.1$, and
$1/(k_{F}a_{s})=0$ (at unitarity), with arbitrary energy unit.}
\label{fig:potential-ex}
\end{figure}

We show in Fig~\ref{fig:potential-ex} a typical contour plot of the
thermodynamic potential $\Omega_0$ as functions of $\Delta$ and
$\vert \mathbf{Q} \vert$, where lighter regions denote lower $\Omega_0$.
Notice that the normal phase occurs at $\Delta = 0$ for all values of
$\vert \mathbf{Q} \vert$. In this plot, two local minima are present,
corresponding to the normal ($\Delta =0$) and the BCS
($\Delta \neq 0$, $\vert \mathbf{Q}\vert$ =0) states, respectively.
Combining the results together with those obtained from the GL theory,
we can get more information about the finite temperature phase diagrams
of trapped Fermi gases, which are discussed next.


\section{Finite temperature phase diagram of trapped Fermi gases with
population imbalance}

\label{sec:phasediagram}

Up to now, we discuss a GL theory which is valid for the small order
parameter regime but not constrained by any approximation schemes, as well
as a mean-field approach which is approximate but works for wider
regions. Using these methods and the LDA approximation, we are able to
analyze the finite temperature phase diagrams of polarized Fermi gases in
harmonic traps. Next, we first consider in Section~\ref{sec:resonance} the
case of unitarity, where the scattering length $a_{s}$ does not set a length
or energy scale, leading to the presence of universality. In Section~\ref{sec:BCS},
we discuss the weakly interacting BCS regime, where emphasis is
put on the possibility of a stable LOFF state. Lastly, the strongly
interacting BEC regime is studied in Section~\ref{sec:BEC}.


\subsection{Unitarity}

\label{sec:resonance}

Following the formalism outlined in the previous section, we first map out
the phase diagram for trapped fermions at unitarity. The different phases in
the trap can be identified from the global minimum of the mean-field
thermodynamic potential at order parameter $\Delta $ and wave vector
$\mathbf{Q}$. At zero temperature, there are three phases which could be
possibly present: (i) a BCS superfluid state with $\Delta \neq 0$ at
$|\mathbf{Q}|=0$; (ii) a normal mixed state (NM) with $\Delta =0$ and two
Fermi surfaces ($\mu _{\uparrow },\mu _{\downarrow }>0$); and (iii) a normal
polarized state (NP) with $\Delta =0$ and one Fermi surface
($\mu _{\uparrow}>0$, $\mu _{\downarrow }<0$). The trap boundary is set
when $\Delta =0$ and both Fermi surfaces vanish
($\mu _{\uparrow },\mu _{\downarrow }<0$). However, at finite temperatures,
both the trap boundary and the phase boundary between NM and NP are blurred.

\begin{figure}[!]
\begin{center}
\psfrag{r}{$r$}
\includegraphics[width=8.0cm]{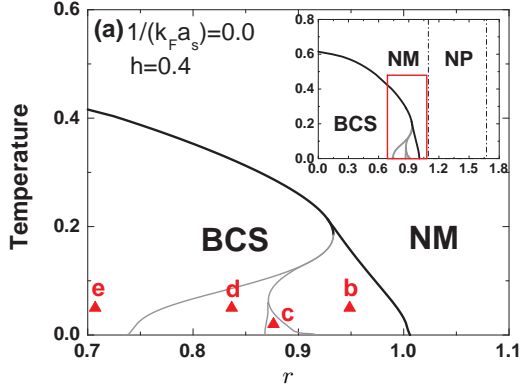}
\psfrag{Q}{$\vert {\bf Q}\vert$}
\psfrag{D}{$\Delta$}
\includegraphics[width=8.0cm]{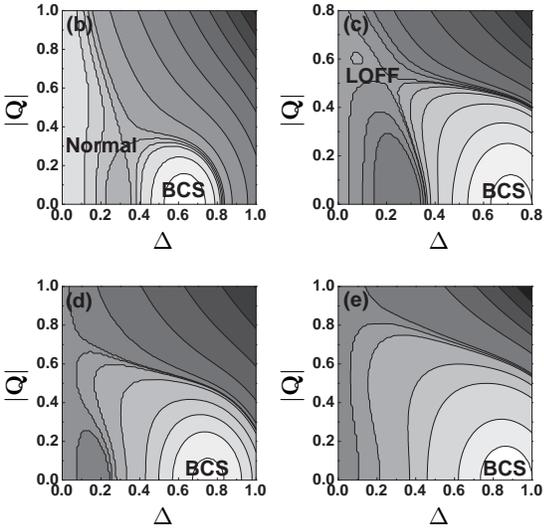}
\end{center}
\caption{(a) Finite temperature phase diagram of a polarized Fermi gas in a
trap with $h=0.4$ and $1/(k_F a_s)=0$, where the chemical potential at the
trap center $\mu_0$ is used as energy unit, and $r_0 \equiv \sqrt{2 \mu_0/(m \omega^2)}$
is used as length unit. The complete phase diagram is shown in the inset,
while the selected area (rectangle) is zoomed in to show detailed structures.
In this plot, a superfluid BCS, a normal mixed (NM), and a normal polarized (NP)
states can be sequentially identified from trap center to trap edge.
While the BCS and normal regions are separated by a phase transition line (dark solid),
the trap boundary and the phase boundary between NM and NP (dot-dashed) are
blurred at finite temperatures. Furthermore, within the BCS phase, four regions
can be identified due to the existence of metastable normal and LOFF states
(gray solid). The contours of thermodynamic potential at representative points
of each region are plotted in (b-e), showing corresponding characteristic
behaviors.}
\label{fig:phase-unitary}
\end{figure}

We show in Fig.~\ref{fig:phase-unitary}(a) the finite temperature phase
diagram for a trapped Fermi gas with population imbalance. From the trap
center to the edge, the BCS, normal mixed (NM), and normal polarized (NP)
phases are identified sequentially. Furthermore, within the BCS phase, four
regions can be identified due to the existence of normal and LOFF metastable
states, which can be explicitly shown from the thermodynamic potential. At
low temperatures, the thermodynamic potential acquire a double well
structure near the trap edge. Starting from the trap edge, the normal
minimum at $\Delta =0$ is lower than the BCS minimum at ($\Delta \neq 0$,
$|\mathbf{Q}|=0$), while the relative order reverses after crossing the
normal-BCS phase boundary [see Fig.~\ref{fig:phase-unitary}(b)]. Continuing
towards the trap center, the normal minimum becomes unstable due to LOFF
instability, but remains stable for BCS instability. The LOFF instability
can lead to a metastable LOFF state at low temperatures, as shown in
Fig.~\ref{fig:phase-unitary}(c). In this case, the global minimum in the
landscape still corresponds to a BCS state. However, from the normal state
to the BCS state, one needs to pass a potential barrier through a first-order
phase transition. Further to the trap center or at higher temperature, the LOFF
state loses its metastability such that the double well structure disappears
and the BCS state becomes the only minimum in the landscape of the
thermodynamic potential, as shown in Fig.~\ref{fig:phase-unitary}(d). What
is special in this case is that to go from a normal state to a BCS state
(with $|\mathbf{Q}|=0$), one needs to follow a path of LOFF instability
with pair momentum $|\mathbf{Q}|\neq 0$ (there is no BCS
instability at $|\mathbf{Q}|=0$ when the order parameter $\Delta $ is
small). As one moves even further towards the trap center, the normal
minimum becomes unstable due to both LOFF and BCS instabilities, as
illustrated in Fig.~\ref{fig:phase-unitary}(e), and the system goes to the
BCS phase directly through the BCS instability at $|\mathbf{Q}|=0$,
corresponding to a second order phase transition.

\begin{figure}[!]
\begin{center}
\psfrag{r}{$r$}
\includegraphics[width=8.0cm]{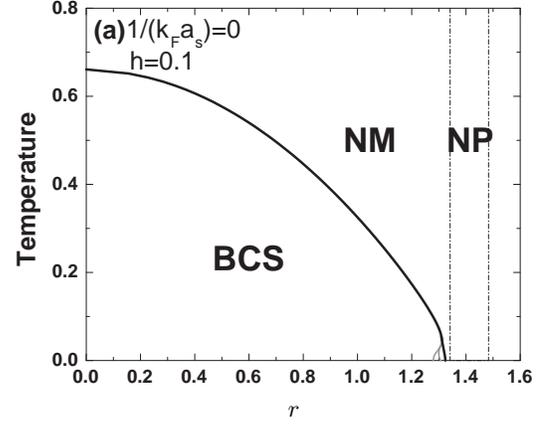}
\includegraphics[width=8.0cm]{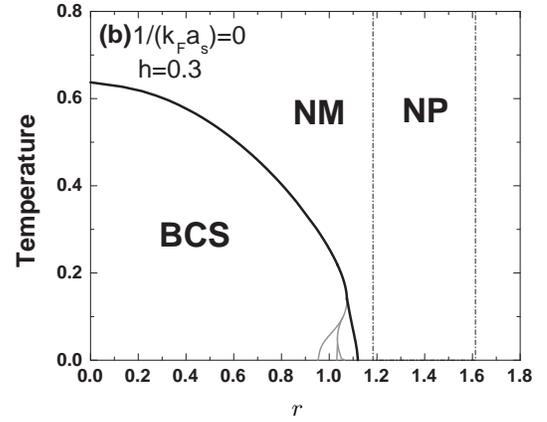}
\includegraphics[width=8.0cm]{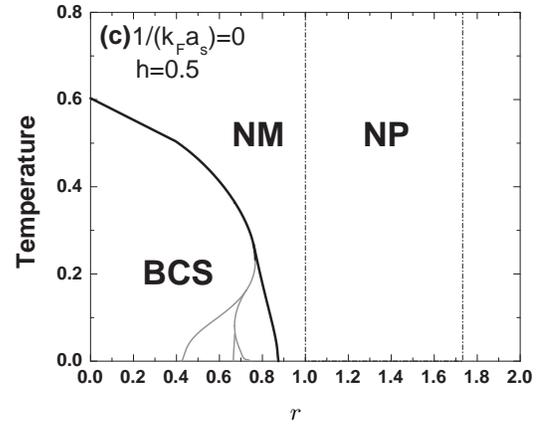}
\end{center}
\caption{Finite temperature phase diagrams of polarized Fermi gases in a
harmonic trap at unitarity $1/(k_F a_s)=0$, where the same energy and length
units are used as in Fig.~\ref{fig:phase-unitary}(a). Parameters used in these
plots are (a) $h=0.1$; (b) $h=0.3$; and (c) $h=0.5$.}
\label{fig:phase-unitary-2}
\end{figure}

Considering the universality present at unitarity, the qualitative features
of phase diagrams are identical for arbitrary values of $h>0$, as shown in
Fig.~\ref{fig:phase-unitary-2} (the temperature and phase boundaries get
somewhat rescaled). Therefore, we can conclude that for a polarized Fermi
gas in a harmonic trap where the interaction is tuned at resonance, a globally
stable LOFF state can not be present on the finite temperature phase diagram,
although there exist a small region of a metastable LOFF state and also a region
where the normal state becomes unstable due to the LOFF instability (but still
ends to a BCS state). This conclusion is consistent with the findings in
Refs.~\cite{sheehy-06} for the zero-temperature case, but does not agree with the
outcomes in~\cite{he-07,machida-06, kinnunen-06}.


\subsection{BCS regime}

\label{sec:BCS}

\begin{figure}[tbp]
\begin{center}
\psfrag{r}{$r$}
\includegraphics[width=8.0cm]{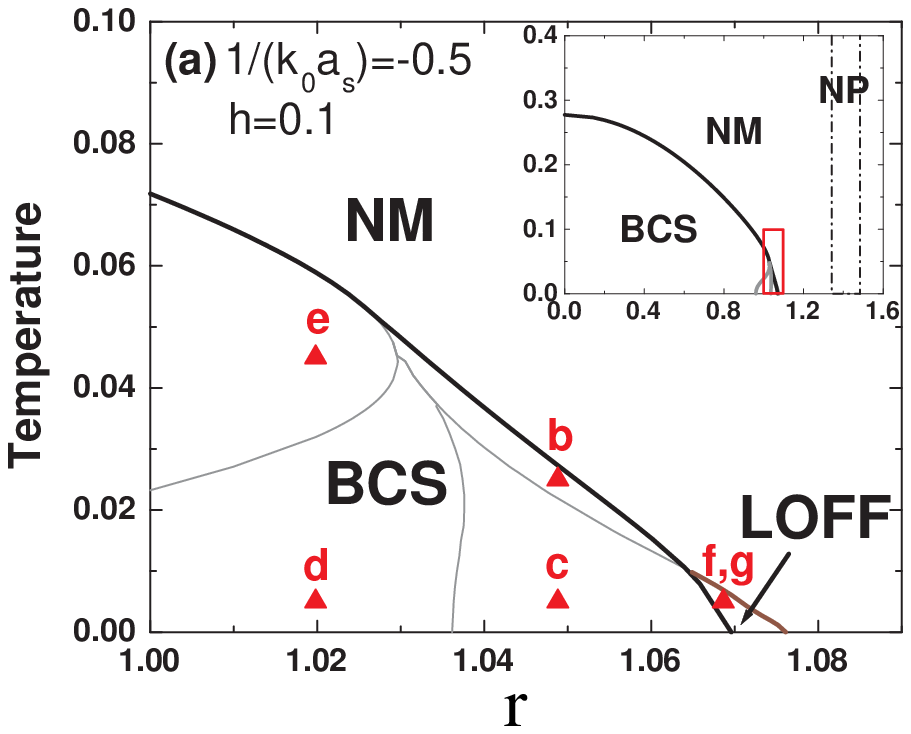}
\psfrag{Q}{$\vert {\bf Q}\vert$}
\psfrag{D}{$\Delta$}
\includegraphics[width=8.0cm]{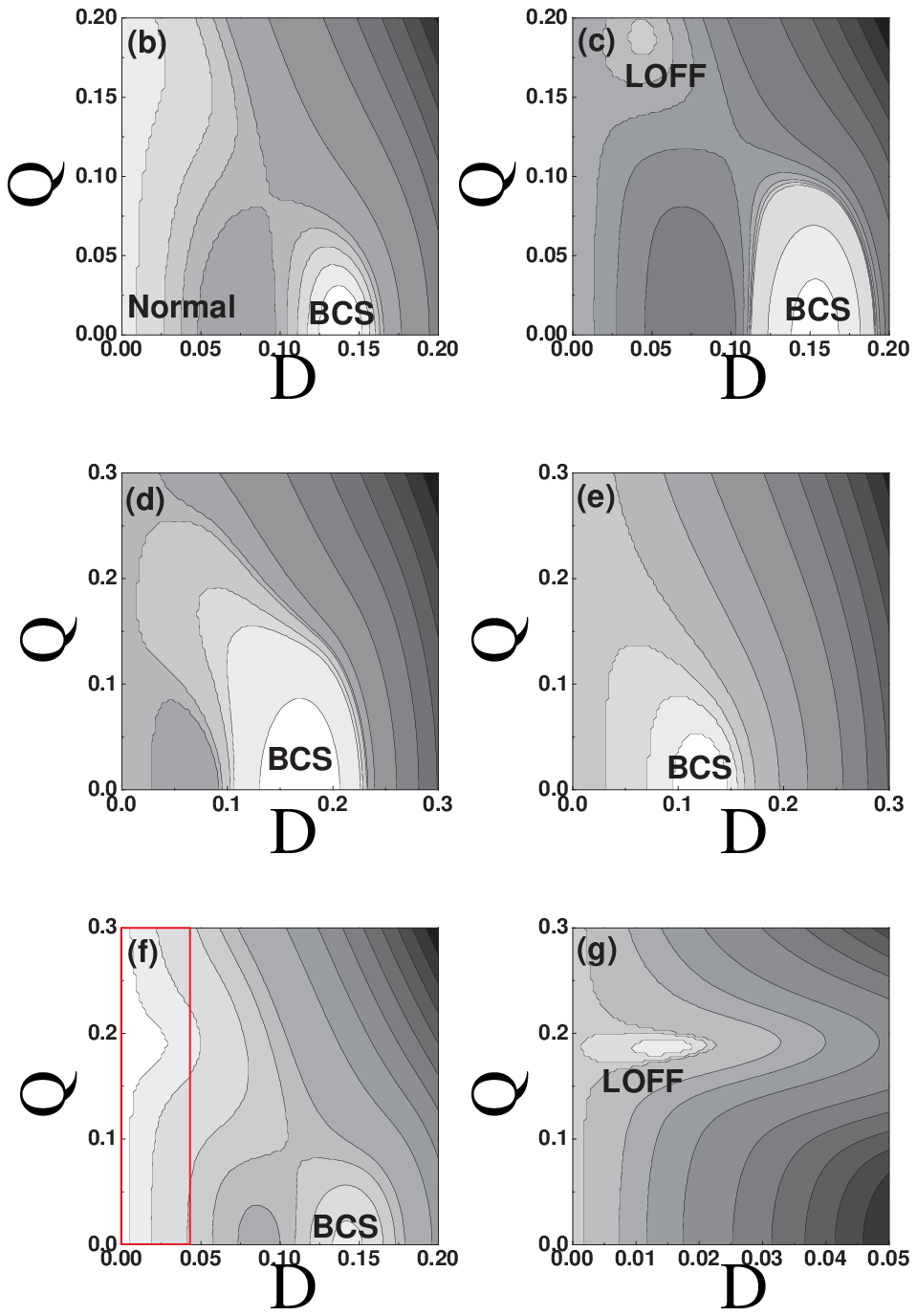}
\end{center}
\caption{(a) Finite temperature phase diagram for a polarized Fermi gas in a
harmonic trap, showing the small region where a stable LOFF state is
present. The complete phase diagram is shown in the inset, while the
selected area (rectangle) is zoomed in to show the detailed structure.
The parameters used in this plot are $h=0.1$ and $1/(k_{0}a_{s})=-0.5$
(BCS regime), where $k_{0}=\sqrt{2m \mu _{0}}$ with $\mu_{0}$ is set
as the energy unit. The radius $r$ is in unit of
$r_0 \equiv \sqrt{2 \mu_0/(m \omega^2)}$.
Within the BCS phase, four regions can be identified due to the stability
of normal and LOFF states and how they are broken (gray solid). The
thermodynamic potential at representative points of each region are
plotted in (b-f), while (g) shows the detailed structure of
the selected area (rectangle) in plot (f).}
\label{fig:phase-BCS}
\end{figure}

Comparing to the unitarity case, the phase diagrams on the BCS side of
Feshbach resonances are more complicated. In Fig.~\ref{fig:phase-BCS}(a),
we show a typical case on the BCS regime. The most significant feature is the
presence of a stable LOFF state at low temperatures, which is characterized
by a global minimum of the thermodynamic potential at $\Delta (\mathbf{Q})\neq 0$
with $|\mathbf{Q}|\neq 0$, as shown in the contour plots
Fig.~\ref{fig:phase-BCS}(f) and (g).

In addition to the presence of a stable LOFF state, the BCS phase also
contains four regions which can be identified due to the existence of
metastable normal and LOFF states. By evaluating the thermodynamic potential
$\Omega_0$, the four regions can be characterized as $\Omega_0$:
(i) has two minima corresponding to a stable BCS and a metastable normal
states [see Fig.~\ref{fig:phase-BCS}(b)];
(ii) has two minima corresponding to a stable BCS
and a metastable LOFF states [see Fig.~\ref{fig:phase-BCS}(c)];
(iii) has only one BCS minimum while the
normal state is stable for BCS instability, but unstable for LOFF
instability [see Fig.~\ref{fig:phase-BCS}(d)];
and (iv) has only one BCS minimum and the normal state is unstable for
BCS instability [see Fig.~\ref{fig:phase-BCS}(e)].
\begin{figure}[!]
\begin{center}
\psfrag{r}{$r$}
\includegraphics[width=7.4cm]{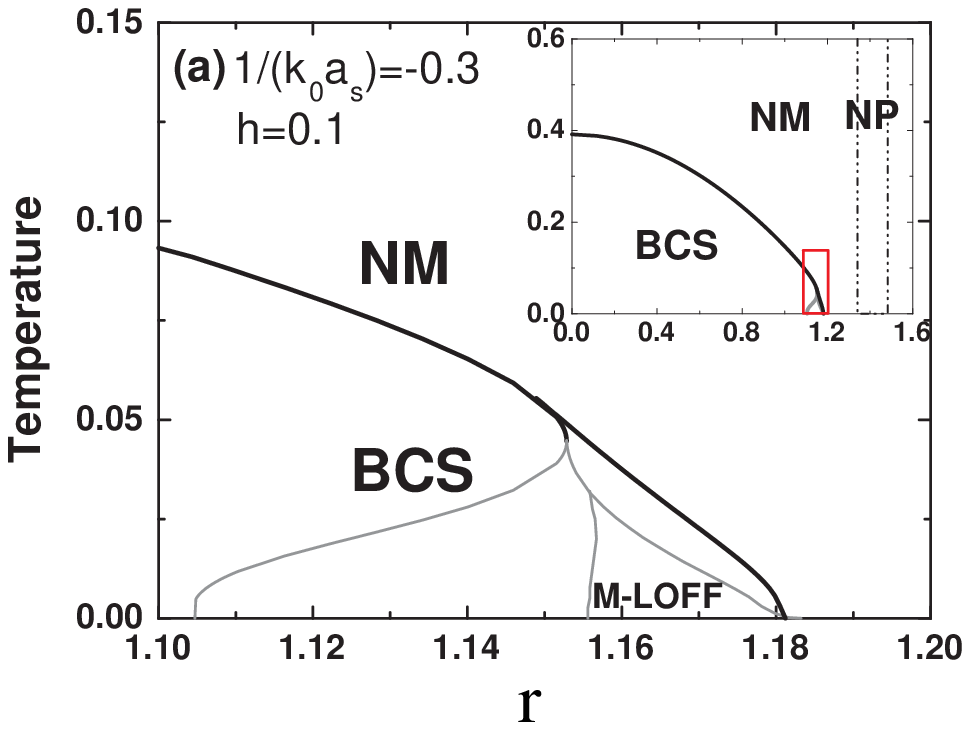}
\includegraphics[width=7.4cm]{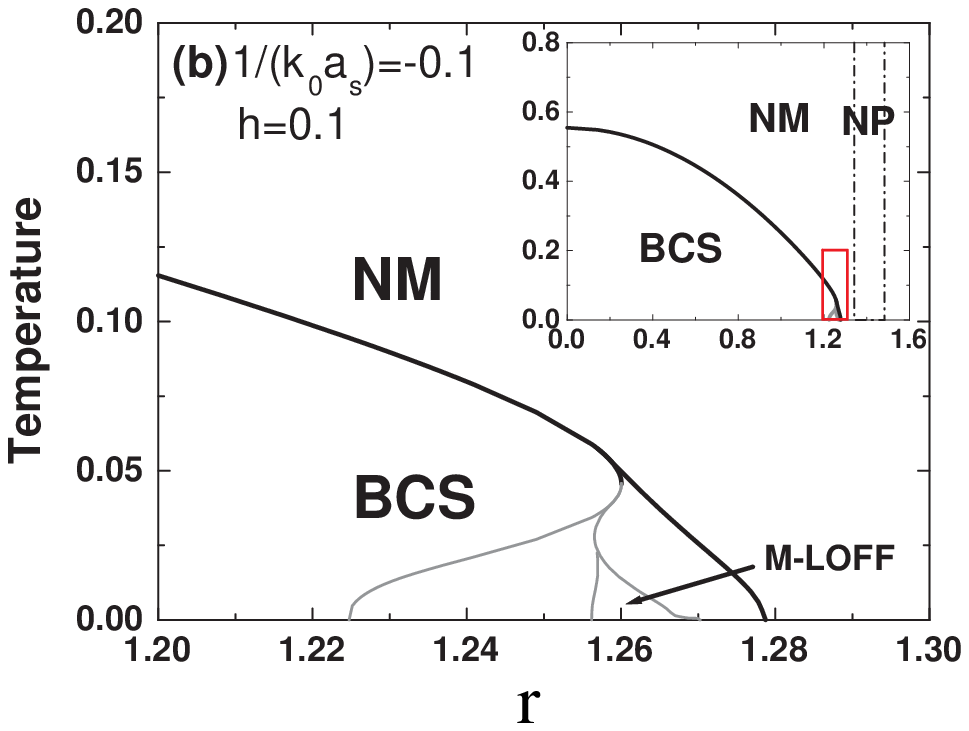}
\includegraphics[width=7.4cm]{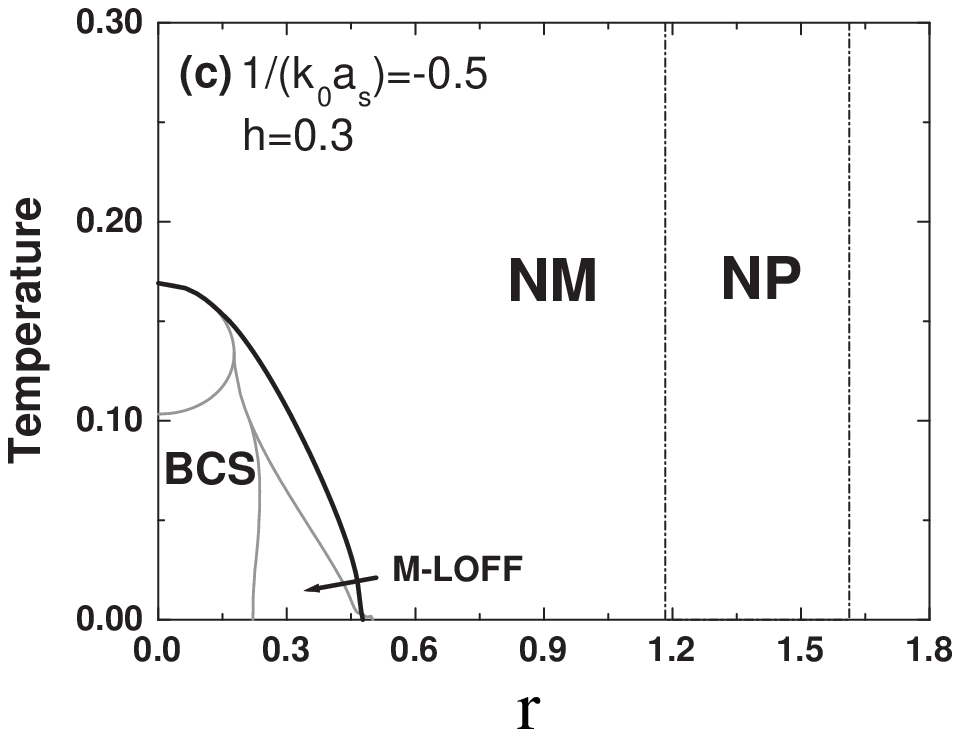}
\end{center}
\caption{Finite temperature phase diagrams of polarized Fermi gases in a
harmonic trap in the BCS regime, with the same energy and length units
as used in Fig.~\ref{fig:phase-BCS}(a).
In (a) and (b), the complete phase diagrams are shown in the insets,
while the selected areas (rectangle) are zoomed in to show detailed structures.
Notice in these two plots that for fixed chemical potential difference $h=0.1$,
the region with stable LOFF state becomes negligible and the region
with metastable LOFF state [M-LOFF, also characterized by point c in
Fig.~\ref{fig:phase-BCS}(a)] shrinks as moving towards unitarity.
Furthermore, if fixing the value of $1/(k_0 a_s)=-0.5$ as in
Fig.~\ref{fig:phase-BCS}, the stable LOFF state also almost disappears
by increasing $h$ to $h=0.3$, as shown in (c). Other parameters used
in these plots are (a) $1/(k_0 a_s) =-0.3$, and (b) $1/(k_0 a_s) =-0.1$,
where $k_0 = \sqrt{2 m \mu_0}$.}
\label{fig:phase-BCS-2}
\end{figure}
\begin{figure}[!]
\begin{center}
\psfrag{r}{$r$}
\includegraphics[width=7.4cm]{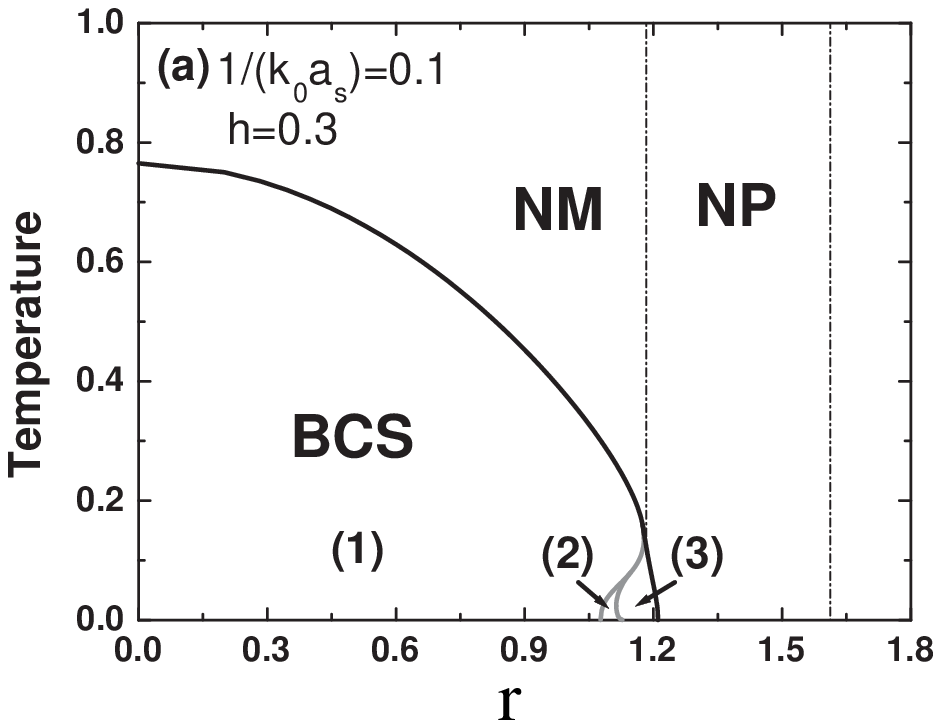}
\includegraphics[width=7.4cm]{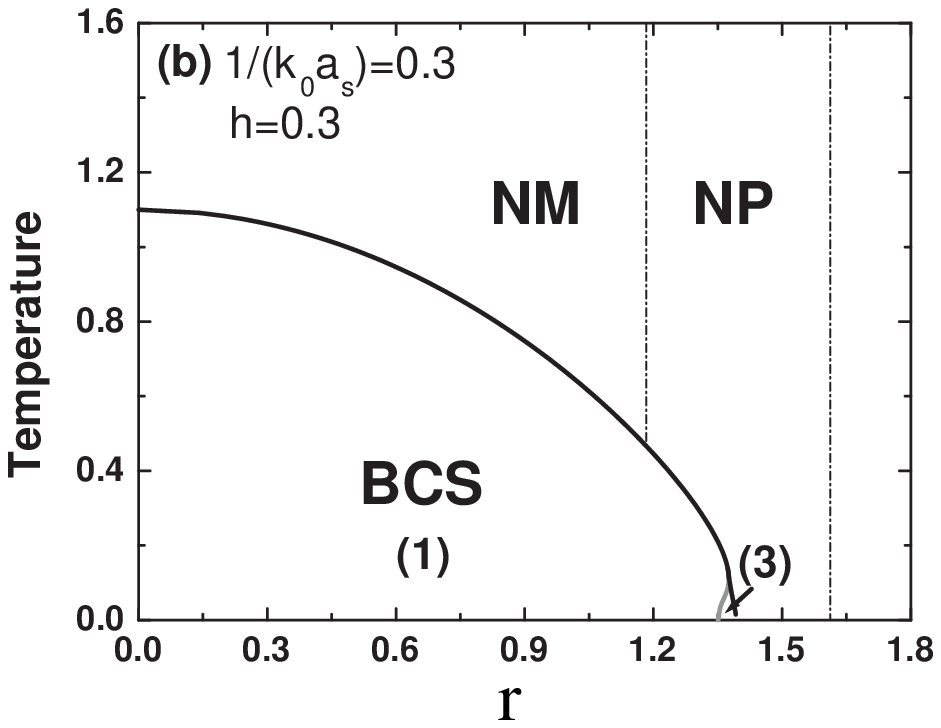}
\includegraphics[width=7.4cm]{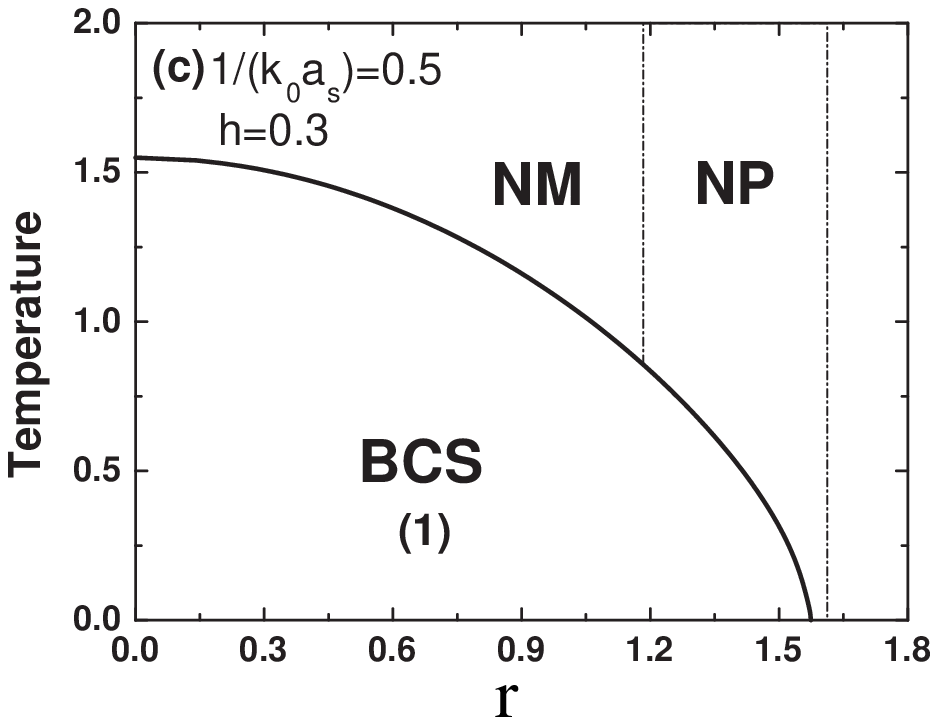}
\end{center}
\caption{Finite temperature phase diagrams of polarized Fermi gases in a
harmonic trap in the BEC regime. The same energy and length units are
used as in Fig.~\ref{fig:phase-BCS-2}.
Three regions may be present in these plots, where the thermodynamical
potential (1) has only one BCS minimum and the normal state is unstable
for BCS and LOFF instabilities; (2) has only one BCS minimum and the
normal state is unstable only for LOFF instability; and (3) has two minima
corresponding to a stable BCS and a metastable normal states.
Parameters used in these plots are
(a) $1/(k_{0}a_{s})=0.1$, $h=0.3$; (b) $1/(k_{0}a_{s})=0.3$, $h=0.3$;
and (c) $1/(k_{0}a_{s})=0.5$, $h=0.3$, where $k_0 = \sqrt{2 m \mu_0}$.}
\label{fig:phase-BEC}
\end{figure}

It should be emphasized that unlike the unitarity case, universality is
not present in the BCS regime, since the finite scattering parameter $a_{s}$
sets an additional length or energy scale. Therefore, the characteristics
of phase diagrams for various $1/(k_{0}a_{s})<0$ and $h>0$ are not
identical. However, as shown in a series of phase diagrams in
Fig.~\ref{fig:phase-BCS-2}, we can summarize some common features within a
qualitative level. First, the region where a LOFF state is stable is only
present in the BCS regime, and disappears as one moves towards the
unitarity [see Fig.~\ref{fig:phase-BCS-2}(b)]. Second, a stable LOFF state
is only present in a small region of the trap at low temperatures,
and restricted for smaller chemical potential difference $h$.
By increasing $h$ (or equivalently polarization $P$), the region for
stable LOFF state shrinks as shown in Fig.~\ref{fig:phase-BCS-2}(c).

Although our calculation results can not be directly compared with
experiments, where the total particle number and polarization are
observables instead of $\mu_{0}$ and $h$, the plots in Fig.~\ref{fig:phase-BCS-2}
outline the general features of a finite temperature phase
diagram in the BCS regime. Therefore, we can conclude that a stable LOFF
state may be present in harmonically trapped Fermi gases with population
imbalance, but only at low temperatures and in a small region of the trap.


\subsection{BEC regime}

\label{sec:BEC}

As concluded in the previous discussion, the LOFF state becomes less
favorable as one moves from the BCS regime to the unitarity. As one further
increases the interaction strength, the same trend is kept in the BEC
regime, as shown in Fig.~\ref{fig:phase-BEC}. Although the finite scattering
length $a_{s}$ sets an additional energy scale such that the phase diagrams
are qualitatively different for various $1/(k_{0}a_{s})>0$ and $h>0$, there
are still some general features which can be concluded. First, similar to
the unitarity case, a stable LOFF state is not present on the finite
temperature phase diagrams in the BEC regime. Second, the structure of the
BCS phase becomes simpler and contains less sub-regions.
By moving from the unitarity towards the BEC limit,
the region with metastable LOFF state disappears first
such that at the value of $1/(k_0 a_s)=0.1$ [see Fig.~\ref{fig:phase-BEC}(a)],
only three regions are present in the BCS phase.
By increasing the value of $1/(k_0 a_s)$, the region where
the normal state becomes unstable only due to LOFF instability
[region (2) in Fig.~\ref{fig:phase-BEC}(a)] disappears,
such that the BCS phase becomes even simpler as shown in
Fig.~\ref{fig:phase-BEC}(b).
Further towards the BEC side, only a simple BCS superfluid state
[region (1)] is present at the trap center, leading to a phase diagram
similar to the unpolarized case [see Fig.~\ref{fig:phase-BEC}(c)].
Notice that for the parameters used in these plots,
the breached pair state (BP1)~\cite{liu-03,wei3} characterized by
gapless fermionic excitations with a Fermi surface
is not present at zero temperature. At finite temperatures,
the boundary between BCS and BP1 is blurred and the crossover
cannot be clearly distinguished.


\section{Summary}

\label{sec:conclusions}

We discuss in this manuscript the finite temperature phase diagrams
of a trapped Fermi gas with population imbalance, focusing on the existence
and stability of the LOFF state. We first derive a Ginzburg-Landau (GL)
theory to study the second order normal-BCS and normal-LOFF phase
transitions, where the order parameter is assumed to be small to ensure
validity. Furthermore, in order to determine the complete phase diagram, we
adopt the mean-field approximation and directly minimize the resulting
thermodynamic potential. This method allows us to distinguish the stable,
the metastable, and the unstable saddle point phases from solutions of the
order parameter equations.

Using these methods, we are able to map out the finite-temperature phase
diagrams over the BCS to BEC region. We show that a stable LOFF
state exists only on the BCS side of the Feshbach resonance, but not at
unitarity or in the BEC regime. Furthermore, we find that the LOFF state
only exists at low temperatures within an appropriate region of population
imbalance. With a global harmonic trap, even in the most optimal situation
of all the parameters, the LOFF state is only present in a small region of
the trap. With the experimental limits on temperature and spatial
resolution, we expect that it is very challenging to make a direct
observation of a stable LOFF state in typical harmonic traps.

This work is supported by the MURI, the DARPA, the NSF under
grant number 0431476, the ARDA under ARO contracts,
and the A. P. Sloan Foundation.



\begin{references}
\bibitem{note2}
We assume here the trap has a spherical symmetry without loss
of generality, as any non-spherical harmonic trap
can be rescaled to a spherical one under the LDA~\cite{wei3}.
\end{references}
\end{document}